\documentclass{pinchcr}

\usepackage{amsmath}
\usepackage{amssymb}
\usepackage[english]{babel}
\usepackage[superscript]{cite}
\usepackage{calrsfs}
\usepackage{indentfirst}
\usepackage{mathrsfs}

\begin{document}
	\begin{center}
		Landau Quantized Dynamics and Spectrum of the Diced Lattice \\[4mm]
		N.J.M. Horing \\
		Department of Physics \\
		Stevens Institute of Technology \\
		Hoboken, N.J. 07030, USA\\[2mm]
		July 13, 2020\\[4mm]
		\textbf{Abstract}\\[4mm]
	\end{center}
	
	In this work the role of magnetic Landau quantization in the dynamics and spectrum of Diced Lattice charge carriers is studied in terms of the associated pseudospin 1 Green's function. The equations of motion for the 9 matrix elements of this Green's function are formulated in position/frequency representation and are solved explicitly in terms of a closed form integral representation involving only elementary functions. The latter is subsequently expanded in a Laguerre eigenfunction series whose frequency poles identify the discretized energy spectrum for the Landau-quantized Diced Lattice as $\epsilon_{n}=\pm\sqrt{2(2n+1)\alpha^{2}eB}$ ($\alpha\sqrt{2}$ is the characteristic speed for the Diced Lattice) which differs significantly from the nonrelativistic linear dependence of $\epsilon_{n}$ on $B$, and is similar to the corresponding $\sqrt{B}-$dependence of other Dirac materials (Graphene, Group VI Dichalcogenides).
	
	\section*{1 \hspace{1em} Introduction}
	
	The magnetic field has traditionally played an important role in studies of the physics of materials, as an agent for probing the properties of matter and inducing new physical features $^{1,2}$. Such studies involving the magnetic field in analyses of nonrelativistic carrier dynamics (Fermi surfaces, transport and collective phenomena) date back to the 1950's and continue to the present for three- and two- dimensional systems. Our scientific and engineering establishments are now dealing with relativistic systems, the Dirac materials, $^{3,4}$ whose (low) energy spectra and Hamiltonians mimic those of relativistic electrons/positrons in being proportional to momentum. As such, they attract intense intellectual interest as material "laboratories" realizing relativistic phenomenology. In regard to practical importance, the Dirac materials provide us with candidates to succeed silicon as the material of choice for the next generation of computers and electronic devices (and much else, including medical applications). It started with the discovery of the exceptional electrical conduction and sensing properties of graphene $^{5-8}$ by Geim and Novoselov (for which they received the 2010 Nobel prize), which has become one of the most heavily researched materials of all time. $^{9-11}$ Other materials of the Dirac type have been found, including Silicene $^{12}$, Topological Insulators $^{13}$, Group VI Dichalcogenides $^{14}$, the Diced Lattice $^{15,16}$ etc. and the literature is now filling with further research reports on them.
	
	In this paper we address the fundamental dynamics and spectrum of the pseudospin 1 Diced Lattice subject to a normal magnetic field. As a two-dimensional system, the spectrum is fully discretized by Landau quantization induced by the applied field. In Section 2 we formulate the 9 equations of motion for the 9 elements of the Green's function of the Diced Lattice in position/frequency representation, subject to the magnetic field. These equations are all solved exactly in Section 3 in terms of a simple, closed form integral representation involving only elementary functions. This integral representation is seen to involve the generating function for Laguerre polynomials, and the associated expansion is tantamount to an eigenfunction expansion of the Green's function, with frequency poles that identify the Landau quantized eigenenergies of the Diced Lattice. Having thus obtained the Landau quantized Green's function for the magnetized Diced Lattice we remark on possible applications in Section 4, some of which are in progress.
	
	\section*{2 \hspace{1em} Green's Function Equations for the Diced Lattice in a Normal Quantizing Magnetic Field}
	The Hamiltonian ($H$) of the Diced Lattice in a constant, uniform normal magnetic field \textbf{B} involves the canonical momentum components $\pi_{x}$, $\pi_{y}$,
	\begin{equation}\tag{2.1}
	\pi_x = \frac{1}{i}\frac{\partial}{\partial x} + \frac{eB}{2}y \hspace{1em} ;\hspace{1em}\pi_y = \frac{1}{i}\frac{\partial}{\partial y} - \frac{eB}{2}x,
	\end{equation}
	in the form
	\begin{equation}\tag{2.2}
	\pi_+ = \pi_x + i\pi_y \hspace{1em} ;\hspace{1em}\pi_- = \pi_x - i\pi_y,
	\end{equation}
	such that the 3$\times$3 pseudospin matrix $H$ $^{15,16}$ is given in position/frequency representation as:
	\begin{equation}\tag{2.3}
	H = 
	\begin{bmatrix}
	0 & \alpha\pi_- & 0 \\
	\alpha\pi_+ & 0 & \alpha\pi_-\\
	0 & \alpha\pi_+ & 0
	\end{bmatrix},
	\end{equation}
	where $\alpha = \hslash v/\sqrt{2}$ ($v$ is a characteristic speed of the Diced Lattice, $\hslash\rightarrow1$). The associated pseudospin matrix Green's function equation is given by \\($I$ is the 3$\times$3 unit matrix)
	\begin{equation}\tag{2.4}
	(\omega - I) G(\textbf{r},\textbf{r}^{\prime};\omega) = I \delta^{(2)} (\textbf{r} - \textbf{r}^{\prime}), [\omega\rightarrow \omega+i0^{+}\thinspace\text{for the retarded function}]
	\end{equation}
	where $\textbf{r} = (x, y); \: \textbf{r}^{\prime} = (x^{\prime}, y^{\prime});$ and \textbf{R} = \textbf{r} - $\textbf{r}^{\prime},X=x-x^{\prime},Y=y-y^{\prime}$;\\[1mm]
	
	This Green's function equation is readily simplified by considering gauge invariance, by which the change of gauge from vector potential $\textbf{A} = \frac{1}{2}\textbf{B}\times\textbf{r}$ to $\textbf{A}^{\prime} = \frac{1}{2}\textbf{B}\times(\textbf{r} - \textbf{r}^{\prime}) = \frac{1}{2}\textbf{B}\times\textbf{R}$ yields a translationally invariant equation for the gauge-transformed Green's function counterpart $G^{\prime}(\textbf{R};\omega)$, with $G^{\prime}$ related to $G$ by the Peierls phase factor $C(\textbf{r},\textbf{r}^{\prime})$, $^{17,18}$
	\begin{equation}\tag{2.5}
	G(\textbf{r},\textbf{r}^{\prime};\omega) = C(\textbf{r},\textbf{r}^{\prime})G^{\prime}(\textbf{R};\omega),
	\end{equation}
	where $C(\textbf{r},\textbf{r}^{\prime})$ = $\exp$($\frac{ie}{2}\textbf{r}\cdot\textbf{B}\times\textbf{r}^{\prime}-\phi(\textbf{r})+\phi(\textbf{r}^{\prime})$) and $\phi(\textbf{r})$ is an arbitrary gauge function, which we take to vanish identically. The resulting translationally invariant equation for $G^{\prime}(\textbf{R},\omega)$ is given by
	\begin{equation}\tag{2.6}
	(\omega I-H(\textbf{R},\omega))G^{\prime}(\textbf{R},\omega)=I\delta^{2}(\textbf{R}),
	\end{equation}
	since the change of gauge made the vector potential, and hence the gauge-transformed Hamiltonian, depend only on $\textbf{R}=(\textbf{r}-\textbf{r}^{\prime})$ to the exclusion of any $(\textbf{r}+\textbf{r}^{\prime})$ dependance. Accordingly, $\pi_{x}\rightarrow\pi_{X};\pi_{y}\rightarrow\pi_{Y}$ in $\pi_{\pm}\rightarrow\pi_{X}\pm i\pi_{Y}$ in the equation for $G^{\prime}(\textbf{R};\omega)$.
	
	The resulting equations for the 9 elements of $G^{\prime}(\textbf{R},\omega)$ coming from Eq (2.6) are given by:
	\begin{equation}\tag{2.7}
	\omega G^{\prime}_{11}-\alpha\pi_{-}G^{\prime}_{21}=\delta^{2}(\textbf{R}),
	\end{equation}
	\begin{equation}\tag{2.8}
	-\alpha\pi_{+}G^{\prime}_{11}+\omega G^{\prime}_{21}-\alpha\pi_{-}G^{\prime}_{31}=0,
	\end{equation}
	\begin{equation}\tag{2.9}
	-\alpha\pi_{+}G^{\prime}_{21}+\omega G^{\prime}_{31}=0,
	\end{equation}
	and
	\begin{equation}\tag{2.10}
	\omega G^{\prime}_{12}-\alpha\pi_{-}G^{\prime}_{22}=0,
	\end{equation}
	\begin{equation}\tag{2.11}
	-\alpha\pi_{+}G^{\prime}_{12}+\omega G^{\prime}_{22}-\alpha\pi_{-}G^{\prime}_{32}=\delta^{2}(\textbf{R}),
	\end{equation}
	\begin{equation}\tag{2.12}
	-\alpha\pi_{+}G^{\prime}_{22}+\omega G^{\prime}_{32}=0,
	\end{equation}
	and
	\begin{equation}\tag{2.13}
	\omega G^{\prime}_{13}-\alpha\pi_{-}G^{\prime}_{23}=0,
	\end{equation}
	\begin{equation}\tag{2.14}
	-\alpha\pi_{+}G^{\prime}_{13}+\omega G^{\prime}_{23}-\alpha\pi_{-}G^{\prime}_{33}=0,
	\end{equation}
	\begin{equation}\tag{2.15}
	-\alpha\pi_{+}G^{\prime}_{23}+\omega G^{\prime}_{33}=\delta^{2}(\textbf{R}).
	\end{equation}
	Employing Eq. (2.11) for $G^{\prime}_{22}$ and using Eq.(2.10) and (2.12) to eliminate $G^{\prime}_{12}$ and $G^{\prime}_{32}$ in terms of $G^{\prime}_{22}$, we obtain
	\begin{equation}\tag{2.16}
	\left[-\frac{\alpha^{2}}{\omega}(\pi_{+}\pi_{-}+\pi_{-}\pi_{+})+\omega\right]G^{\prime}_{22}=\left[-2\frac{\alpha^{2}}{\omega}\pi^{2}+\omega\right]G^{\prime}_{22}=\delta^{2}(\textbf{R}).
	\end{equation}
	Noting that
	\begin{equation}\tag{2.17}
	\pi^{2}=\pi^{2}_{X}+\pi^{2}_{Y}=-\left(\frac{\partial^{2}}{\partial X^{2}}+\frac{\partial^{2}}{\partial Y^{2}}\right)+\left(\frac{eB}{2}\right)^{2}(X^{2}+Y^{2})+eBL_{Z},
	\end{equation}
	where $L_{Z}=\frac{1}{i}[Y\frac{\partial}{\partial X}-X\frac{\partial}{\partial Y}]$ is the angular momentum operator ($Z$-component) in relative coordinates, we bear in mind that conservation of angular momentum $^{17,18}$ mandates that $L_{Z}G(\textbf{r},\textbf{r}^{\prime};\omega)=0$ and, since $[L_{Z},C(\textbf{r},\textbf{r}^{\prime})]=0$, we also have $L_{Z}G^{\prime}(\textbf{R};\omega)=0$. Accordingly, Eq.(2.16) may be written as
	\begin{equation}\tag{2.18}
	\left\{\omega+2\frac{\alpha^{2}}{\omega}\left[\frac{\partial^{2}}{\partial X^{2}}+\frac{\partial^{2}}{\partial Y^{2}}-\left(\frac{eB}{2}\right)^{2}(X^{2}+Y^{2})\right]\right\}G^{\prime}_{22}(\textbf{R};\omega)=\delta^{2}(\textbf{R}).
	\end{equation}
	In a similar manner, Eq.(2.8) for $G^{\prime}_{21}$ may be used jointly with Eq.(2.7) and (2.9) to eliminate $G^{\prime}_{11}$ and $G^{\prime}_{31}$ in terms of $G^{\prime}_{21}$, with the result
	\begin{equation}\tag{2.19}
	\left\{\omega+2\frac{\alpha^{2}}{\omega}\left[\frac{\partial^{2}}{\partial X^{2}}+\frac{\partial^{2}}{\partial Y^{2}}-\left(\frac{eB}{2}\right)^{2}(X^{2}+Y^{2})\right]\right\}G^{\prime}_{21}(\textbf{R};\omega)=\frac{\alpha}{\omega}\pi_{+}\delta^{2}(\textbf{R}).
	\end{equation}
	Proceding in the same way to develop an equation for $G^{\prime}_{23}$ using Eq.(2.14) with Eqns.(2.13) and (2.15) to eliminate $G^{\prime}_{13}$ and $G^{\prime}_{33}$ in terms of $G^{\prime}_{23}$, we obtain
	\begin{equation}\tag{2.20}
	\left\{\omega+2\frac{\alpha^{2}}{\omega}\left[\frac{\partial^{2}}{\partial X^{2}}+\frac{\partial^{2}}{\partial Y^{2}}-\left(\frac{eB}{2}\right)^{2}(X^{2}+Y^{2})\right]\right\}G^{\prime}_{23}(\textbf{R};\omega)=\frac{\alpha}{\omega}\pi_{-}\delta^{2}(\textbf{R}),
	\end{equation}
	so it is clear that 
	\begin{equation}\tag{2.21}
	G^{\prime}_{23}=G^{\prime*}_{21}.
	\end{equation}
	
	Furthermore, hermiticity of the Hamiltonian and consequently the Green's function provides that $G_{ij}(\textbf{r},\textbf{r}^{\prime};\omega)=G^{*}_{ji}(\textbf{r}^{\prime},\textbf{r};\omega)$, and since $G^{\prime}(\textbf{R})$ is an even function of $\textbf{R}$ and $(C(\textbf{r},\textbf{r}^{\prime}))^{+}=C^{*}(\textbf{r}^{\prime};\textbf{r})=C(\textbf{r},\textbf{r}^{\prime})$, we have
	\begin{equation}\tag{2.22}
	G^{\prime}_{ij}(\textbf{R};\omega)=G^{\prime*}_{ji}(\textbf{R};\omega),
	\end{equation}
	so the diagonal elements are real and corresponding off-diagonal elements across the diagonal are complex conjugates. In the next section we will solve Eq.(2.18) for $G^{\prime}_{22}$, Eq.(2.19) for $G^{\prime}_{21}$ and Eq.(2.20) yields $G^{\prime}_{23}=G^{\prime*}_{21}$. By hemiticity, we then obtain $G^{\prime}_{32}=G^{\prime*}_{23}$ and $G^{\prime}_{12}=G^{\prime*}_{21}$. Noting that the Hamiltonian matrix of Eq.(2.3) has yet another symmetry across the \textit{anti}-diagonal, the Green's function has this symmetry as well, so that,
	\begin{equation}\tag{2.23}
	G^{\prime}_{33}=G^{\prime}_{11}.
	\end{equation}
	Moreover, summation of Eqns.(2.7), (2.11), and (2.15) yields $G^{\prime}_{11}$ and $G^{\prime}_{33}$ as
	\begin{equation}\tag{2.24}
	G^{\prime}_{11}(\textbf{R};\omega)=G^{\prime}_{33}(\textbf{R};\omega)=\frac{1}{2}G^{\prime}_{22}(\textbf{R},\omega)+\frac{1}{2\omega}\delta^{2}(\textbf{R}).
	\end{equation}
	Finally, the last two elements of $G^{\prime}$, namely $G^{\prime}_{13}$ and $G^{\prime}_{31}$, may be obtained from Eq.(2.13) and (2.9) as,
	\begin{equation}\tag{2.25}
	G^{\prime}_{13}=\frac{\alpha}{\omega}\pi_{-}G^{\prime}_{23},
	\end{equation}
	\begin{equation}\tag{2.26}
	G^{\prime}_{31}=\frac{\alpha}{\omega}\pi_{+}G^{\prime}_{21}=G^{\prime*}_{13},
	\end{equation}
	so that the equations for all 9 elements of the Green's function are all identified.
	
	\section*{3 \hspace{1em} Solution of the Green's Function Equations}
	Introducing the operator notation $\Lambda_{\textbf{R}}$,
	\begin{equation}\tag{3.1}
	\Lambda_{\textbf{R}}\equiv\omega+\frac{2\alpha^{2}}{\omega}\left[\frac{\partial^{2}}{\partial X^{2}}+\frac{\partial^{2}}{\partial Y^{2}}-\left(\frac{eB}{2}\right)^{2}(X^{2}+Y^{2})\right],
	\end{equation}
	Eq.(2.18) takes the form
	\begin{equation}\tag{3.2}
	\Lambda_{\textbf{R}}G^{\prime}_{22}(\textbf{R};\omega)=\delta^{2}(\textbf{R});
	\end{equation}
	and we note that the solution of this translationally invariant equation was previously obtained in a closed-form integral representation as $^{17,18}$
	\begin{equation}\tag{3.3}
	G^{\prime}_{22}(\textbf{R};\omega)=-\frac{eB}{4\pi}\int_{0}^{\infty}d\tau\frac{e^{i\omega\tau}}{\sin(\frac{2eB\alpha^{2}\tau}{\omega})}\exp\left(\frac{ieB(X^{2}+Y^{2})}{4\tan(\frac{2eB\alpha^{2}\tau}{\omega})}\right).
	\end{equation}
	
	Addressing Eq.(2.20) for $G^{\prime}_{23}$, we have
	\begin{equation}\tag{3.4}
	\Lambda_{\textbf{R}}G^{\prime}_{23}(\textbf{R};\omega)=\frac{\alpha}{\omega}\pi_{-}\delta^{2}(\textbf{R}).
	\end{equation}
	To solve this, we need the actual Green's function, $\mathscr{G}(\textbf{R},\omega)$, inverse to the operator $\Lambda_{\textbf{R}}$, which obeys the equation
	\begin{equation}\tag{3.5}
	\Lambda_{\textbf{R}}\mathscr{G}(\textbf{R},\textbf{R}^{\prime};\omega)=\delta^{2}(\textbf{R}-\textbf{R}^{\prime}).
	\end{equation}
	Bearing in mind that Eq.(3.2) arises as one element of a gauge-transformed counterpart of the original Green's function equation for $G_{22}(\textbf{r},\textbf{r}^{\prime};\omega)$, we recall that in an earlier nonrelativistic study $^{18}$ involving an operator having the same differential structure as $\Lambda_{\textbf{R}}$ of Eq.(3.1), the original, equivalent equation \textit{prior} to our gauge transformation was given by
	\begin{equation}\tag{3.6}
	\Lambda_{\textbf{R}}[C(\textbf{R},\textbf{R}^{\prime})G^{\prime}(\textbf{R}-\textbf{R}^{\prime};\omega]=\delta^{2}(\textbf{R}-\textbf{R}^{\prime})\hspace{.9em};\hspace{.9em}C(\textbf{R},\textbf{R}^{\prime})=\exp\left(\frac{ie}{2}\textbf{R}\cdot\textbf{B}\times\textbf{R}^{\prime}\right).
	\end{equation}
	This process of inversion of the gauge transformation permits us to identify $\mathscr{G}(\textbf{R},\textbf{R}^{\prime};\omega)$ as
	\begin{equation}\tag{3.7}
	\mathscr{G}(\textbf{R},\textbf{R}^{\prime};\omega)=C(\textbf{R},\textbf{R}^{\prime})G^{\prime}_{22}(\textbf{R}-\textbf{R}^{\prime}),
	\end{equation}
	by comparison of Eq.(3.5) and Eq.(3.6). With this determination of $\mathscr{G}(\textbf{R},\textbf{R}^{\prime};\omega)$, the solution of Eq.(3.4) is given by
	\begin{equation}\tag{3.8}
	G^{\prime}_{23}(\textbf{R};\omega)=\frac{\alpha}{\omega}\int d^{2}\textbf{R}^{\prime}C(\textbf{R},\textbf{R}^{\prime})G^{\prime}_{22}(\textbf{R}-\textbf{R}^{\prime};\omega)\pi_{-}\delta^{2}(\textbf{R}^{\prime}).
	\end{equation}
	Here, $\pi_{-}=\frac{1}{i}\frac{\partial}{\partial X^{\prime}}-\frac{\partial}{\partial Y^{\prime}}+\frac{eB}{2}(Y^{\prime}+iX^{\prime})$, and integration by parts yields
	\begin{align*}\tag{3.9}
	G^{\prime}_{23}(\textbf{R};\omega)&=\frac{\alpha}{\omega}\left\{\left(i\frac{\partial}{\partial X^{\prime}}+\frac{\partial}{\partial Y^{\prime}}\right)\left(C(\textbf{R},\textbf{R}^{\prime})G^{\prime}_{22}(\textbf{R}-\textbf{R}^{\prime};\omega)\right)\right\}_{\textbf{R}^{\prime}=0} \\&=\frac{\alpha}{\omega}\left\{\left(i\frac{eB}{2}[iY-X]\right)G^{\prime}_{22}(\textbf{R};\omega) \right.
	\\& \left. +\left(i\frac{\partial}{\partial X^{\prime}}+\frac{\partial}{\partial Y^{\prime}}\right)G^{\prime}_{22}(\textbf{R}-\textbf{R}^{\prime};\omega)\right\}_{\textbf{R}^{\prime}=0}.
	\end{align*}
	
	To express $G^{\prime}_{22}(\textbf{R};\omega)$ in terms of its frequency poles, we note that its $\tau$-integrand is the generator of Laguerre polynomials $^{19}$, such that $(\phi\equiv2eB\frac{\alpha^{2}}{\omega})$
	\begin{equation}\tag{3.10}
	G^{\prime}_{22}(\textbf{R};\omega)=-\frac{ieB}{2\pi}e^{-eB\frac{R^{2}}{4}}\sum_{n=0}^{\infty}L_{n}\left(\frac{eBR^{2}}{2}\right)\int_{0}^{\infty}d\tau e^{i(\omega-\phi)\tau} e^{-i2n\phi\tau},
	\end{equation}
	and integrating to get the retarded Green's function $(\omega\rightarrow\omega+i0^{+})$ we obtain,
	\begin{equation}\tag{3.11}
	G^{\prime}_{22}(\textbf{R};\omega)=\frac{eB}{2\pi}\omega e^{-eB\frac{R^{2}}{4}}\sum_{n=0}^{\infty}L_{n}\left(\frac{eBR^{2}}{2}\right)\frac{1}{\omega^{2}-2(2n+1)\alpha^{2}eB}.
	\end{equation}
	From the frequency poles we may identify the Landau energy eigenvalues $\epsilon_{n}$ of the Diced lattice carriers as $(\hbar\rightarrow 1)$
	\begin{equation}\tag{3.12}
	\epsilon_{n}\equiv\omega_{n}=\pm\sqrt{2(2n+1)\alpha^{2}eB} \hspace{2em} (n=0\dots\infty).
	\end{equation}
	Furthermore, the off-diagonal element $G^{\prime}_{23}(\textbf{R},\omega)$ is given by Eq.(3.9) as
	\begin{equation}\tag{3.13}
	G^{\prime}_{23}(\textbf{R};\omega)=\frac{\alpha(eB)^{2}}{2\pi}e^{-eB\frac{R^{2}}{4}}\left[iX+Y\right]\sum_{n=1}^{\infty}\frac{L^{1}_{n-1}(eB\frac{R^{2}}{2})}{\omega^{2}-2(2n+1)\alpha^{2}eB}.
	\end{equation}
	Moreover, using Eqns.(2.21, 2.26) so that $G^{\prime}_{31}=\frac{\alpha}{\omega}\pi_{+}G^{\prime*}_{23}$, we have
	\begin{equation}\tag{3.14}
	G^{\prime}_{31}(\textbf{R};\omega)=\frac{\alpha^{2}(eB)^{3}}{2\pi\omega}e^{-eB\frac{R^{2}}{4}}\left[X+iY\right]^{2}\sum_{n=2}^{\infty}\frac{L^{2}_{n-2}(eB\frac{R^{2}}{2})}{\omega^{2}-2(2n+1)\alpha^{2}eB}.
	\end{equation}
	
	These results, Eqns.(3.11, 3.13, 3.14) for $G^{\prime}_{22}, G^{\prime}_{23}, G^{\prime}_{31}$, respectively, suffice to determine all 9 elements of $G^{\prime}$ since we also have the relation of Eq.(2.24) stating that $G^{\prime}_{11}=G^{\prime}_{33}=(\frac{1}{2})G^{\prime}_{22}+(1/2\omega)\delta^{2}(\textbf{R});$ and Eq.(2.21) stating that $G^{\prime}_{21}=G^{\prime*}_{23};$ as well as $G^{\prime}_{32}=G^{\prime*}_{23},G^{\prime}_{13}=G^{\prime*}_{31},G^{\prime}_{23}=G^{\prime*}_{32},G^{\prime}_{12}=G^{\prime*}_{21}$ so $G^{\prime}_{12}=G^{\prime}_{32}$. In connection with these relations it should be noted that the diagonal elements, $G^{\prime}_{22},G^{\prime}_{11}$ and $G^{\prime}_{33}$, are real for real $\omega$, while the off-diagonals are non-real complex conjugates across the diagonal.
	
	\section*{4 \hspace{1em} Remarks}
	
	The explicit analytic results for the Green's function elements (above) clearly exhibit the role of magnetic Landau quantization in the dynamics and spectrum of Diced Lattice charge carriers. The Landau quantized Diced Lattice eigenenergies are given by Eq.(3.12) ($\hbar \rightarrow 1$),
	\begin{equation}\tag{4.1}
	\epsilon_{n}=\pm\sqrt{2(2n+1)\alpha^{2}eB},
	\end{equation}
	which differs significantly from the nonrelativistic linear dependence of $\epsilon_{n}$ proportional to $B$; also, this result is similar to the corresponding square root dependence on $B$ of other Dirac materials, Graphene and the Group VI Dichalcogenides \\($\epsilon_{n}=E_{S_{z}}\pm$$\sqrt{g^{2}+(2n+1\mp1)\gamma^{2}eB}$ for the latter) $^{17}$. This spectrum and its associated degeneracy embodied in the structure of the Green's function are of central importance in the determination of the statistical thermodynamic features of the Diced Lattice, including magnetic oscillations of the de Haas- van Alphen type, similar to those of the dichalcogenides at low temperatures $^{20}$, as well as the specific heat; all of which are helpful in characterizing the material. Finally, this Green's function's explicit description of the dynamical propagation of Diced Lattice charge carriers subject to Landau quantization may be employed to determine the RPA "Ring diagram" polarization function of the system and its magnetoplasmons, also the magnetoconductivity tensor and its magnetopolariton spectrum. It also maybe employed to examine the role of the magnetic field in the quantum dynamics and the thermodynamics of nanostructure features embedded in the Diced Lattice, such as quantum dots, quantum wires, quantum antidot superlattices; as well as providing the basis for analysis of other correlation/interaction problems.
	
	\section*{\hspace{1em} References}
	\noindent
	1.	D. Lei, "Matter in Strong Magnetic Fields," \textit{Rev. Mod. Phys.} \textbf{73}, 629    (2001).\\
	2.	National Research Council, "High Magnetic Field Science and Its Applications in the United States: Current Status and Future Directions", Washington DC, National Academic Press (2013).\\
	3.	J. Wang, S. Deng, Zhongfan Liu and Zhirong Liu, "The Rare 2D Materials with Dirac Cones," National Science Review 2:22-39 (2015).\\
	4.	T.O. Wehling, A.M. Black-Shaffer and A.V. Balatsky, "Dirac Materials," arXiv: 1405.5774 v1 [cond-mat.mtrl-sci] (22 May 2014).\\
	5.	M. Katsnelson, "Graphene: Carbon in Two Dimensions," Cambridge University Press (2012).\\
	6.	H. Aoki and M.S. Dresselhaus, "Physics of Graphene," Springer (2013).\\
	7.	E.L. Wolf, "Graphene: A New Paradigm in Condensed Matter and Device Physics," (2013).\\
	8.	N.J.M. Horing, "Aspects of the Theory of Graphene," Transactions Royal Society \textbf{A 368}, 5525-56 (2010).\\
	9.	J.H. Warner, F. Sch\"affel, A. Bachmatiuk and M.H. Rummelli, "Graphene: Fundamentals and Emergent Applications", Elsevier, 2013.\\
	10.	L.E.F. Foa Torres, S. Roche and J.S. Charlie, "Introduction to Graphene-Based Nanomaterials", Cambridge Univ. Press, 2014.\\
	11.	P.A.D. Gonsalves and N.M.R. Peres, "Introduction to Graphene Plasmonics", World Scientific, 2016.\\
	12.	M.J.S. Spencer, "Silicine," Springer (2016).\\
	13.	S.Q. Shen, "Topological Insulators," Springer (2012).\\
	14.	G.K. Ahluwalia, Editor: "Applications of Chalcogenides: S, Se, Te," Springer (2017).\\
	15. J.D. Malcolm and E.J. Nicol, Phys. Rev. B 93,  165433 (2016).\\
	16. D. Bercioux, D.F. Urban, H. Grabert and W. H\"ausler, Phys. Rev. A 80, 063683 (2009).\\
	17. N.J.M. Horing, "Landau Quantized Dynamics and Spectra for Group VI Dichalcogenides, Including a Model Quantum Wire," AIP Advances 7, 065316 (2017).\\
	18. N.J. Horing; Annals of Physics (NY) 31, 1-63 (1965), Section II.\\
	19. A. Erdelyi, etal., Eds., "Higher Transcendental Functions II," p.189 \#17.\\
	20.	N.J.M. Horing, J.D. Mancini, "Thermal and Magnetic Properties of Landau Quantized Group VI Dichalcogenide Carriers in the Approach to the Degenerate Limit", J. Phys. Comm., IOP, \textit{https://doi.org/10.1088/2399-6528/ab7285} (2020).\\
\end{document}